\documentclass[twocolumn,showpacs,superscriptaddress]{revtex4}

\usepackage{graphicx}
\usepackage{bm}
\begin{document}
\title{Characterization of single-molecule pentanedithiol junctions by
    inelastic electron tunneling spectroscopy and first-principles
    calculations}

\author{Carlos R. Arroyo}
\affiliation{Departamento de F\'{\i}sica de la Materia Condensada,
	Universidad Aut\'{o}noma de Madrid, 28049 Madrid, Spain}
\author{Thomas Frederiksen}
\affiliation{Donostia International Physics Center (DIPC), Paseo Manuel
	de Lardizabal 4, Donostia-San Sebasti\'an, Spain}
\author{Gabino Rubio-Bollinger}
\affiliation{Departamento de F\'{\i}sica de la Materia Condensada,
	Universidad Aut\'{o}noma de Madrid, 28049 Madrid, Spain}
\author{Marisela V\'{e}lez}
\affiliation{Departamento de F\'{\i}sica de la Materia Condensada,
	Universidad Aut\'{o}noma de Madrid, 28049 Madrid, Spain}
\affiliation{Instituto Madrile\~{n}o de Estudios Avanzados en Nanociencia
	(IMDEA-Nanociencia),Campus Cantoblanco, 28049 Madrid, Spain}
\author{Andr\'{e}s Arnau}
\affiliation{Donostia International Physics Center (DIPC), Paseo Manuel
	de Lardizabal 4, Donostia-San Sebasti\'an, Spain}
\affiliation{Centro de F\'isica de Materiales CFM-MPC, Centro Mixto CSIC-UPV, Apdo. 1072,
Donostia-San Sebasti\'an, Spain}
\affiliation{Depto. de F\'isica de Materiales UPV/EHU, Facultad de Qu\'imica, Apdo. 1072,
Donostia-San Sebasti\'an, Spain}
\author{Daniel S\'{a}nchez-Portal}
\affiliation{Centro de F\'isica de Materiales CFM-MPC, Centro Mixto CSIC-UPV, Apdo. 1072,
Donostia-San Sebasti\'an, Spain}
\affiliation{Depto. de F\'isica de Materiales UPV/EHU, Facultad de Qu\'imica, Apdo. 1072,
Donostia-San Sebasti\'an, Spain}
\author{Nicol\'{a}s Agra\"{\i}t}
\affiliation{Departamento de F\'{\i}sica de la Materia Condensada,
	Universidad Aut\'{o}noma de Madrid, 28049 Madrid, Spain}
\affiliation{Instituto Madrile\~{n}o de Estudios Avanzados en Nanociencia
	(IMDEA-Nanociencia),Campus Cantoblanco, 28049 Madrid, Spain}

\date{\today}

\begin{abstract}
We study pentanedithiol molecular junctions formed by means of the
break-junction technique with a scanning tunneling microscope at low
temperatures. Using inelastic electron tunneling spectroscopy and
first-principles calculations, the response of the junction to
elastic deformation is examined. We show that this procedure makes a
detailed characterization of the molecular junction possible. In particular, our results
indicate that tunneling takes place through just a single molecule.
\end{abstract}


\maketitle

One of the most challenging aspects to reach ultimate miniaturization of
electronic devices is the characterization of charge
transport through individual molecules in contact with metallic electrodes
\cite{ratner2005nature}. However, from an experimental point of view, it is very
difficult to ensure that the junction consists of just a single molecule, and---even
when such junctions are realized---it is hard to know the
microscopic arrangement, e.g., how the molecule is bound to the electrodes or the
pathway followed by the electrons in the molecule.
A powerful tool for understanding electron
transport in nanoscopic devices is inelastic electron tunneling spectroscopy (IETS)
\cite{stipe1998,agrait2002prl,smit2002nature}. This technique is
based on the variations in the conductance caused by the excitation of
vibrations by the traversing electrons. Some of the characteristic
molecular vibrations have been identified in junctions comprising a relatively
large number of molecules in self-assembled monolayers of alkanedithiol and other thiolated
molecules \cite{wang2004nl,kushmerick2004nl,okabayashi2008prl}. In junctions of short molecules,
like water or benzene, weakly bound to the electrodes, inelastic spectroscopy has
been used to explore the molecular conformations in a single-molecule junction
\cite{kiguchi2008prl}.
Recently, Hihath \textit{et al.}~\cite{hihath2008nl} measured IETS spectra of junctions with 
short alkanedithiols (propanedithiol) and showed that the average of spectra taken on the same 
conductance plateau were sensitive to changes in the configuration of the junction as a whole 
and were associated with a plastic deformation of the junction.
However, the observed changes  permitted neither to ensure the presence of a single molecule  
at the junction nor to discern its bonding configuration and state of strain, 
something that 
requires a detailed comparison between individual spectra and first-principles calculations 
of the inelastic electronic transport process.

Here, we report the evolution of IETS in single pentanedithiol molecules covalently
bound to gold electrodes as the molecular junction is deformed
\emph{elastically}, i.e., with no change in molecular configuration.
The measurements are performed using a scanning tunneling microscope (STM)
at low temperatures (4.2 K) and compared to first-principles calculations, 
which permits a unique identification of the main peaks in the vibrational
spectra, as well as their variation with strain. This further allows us to conclude unambiguously
that electrons tunnel through just a single molecule suspended between the electrodes.

\begin{figure}[h!]
\includegraphics[width=8 cm]{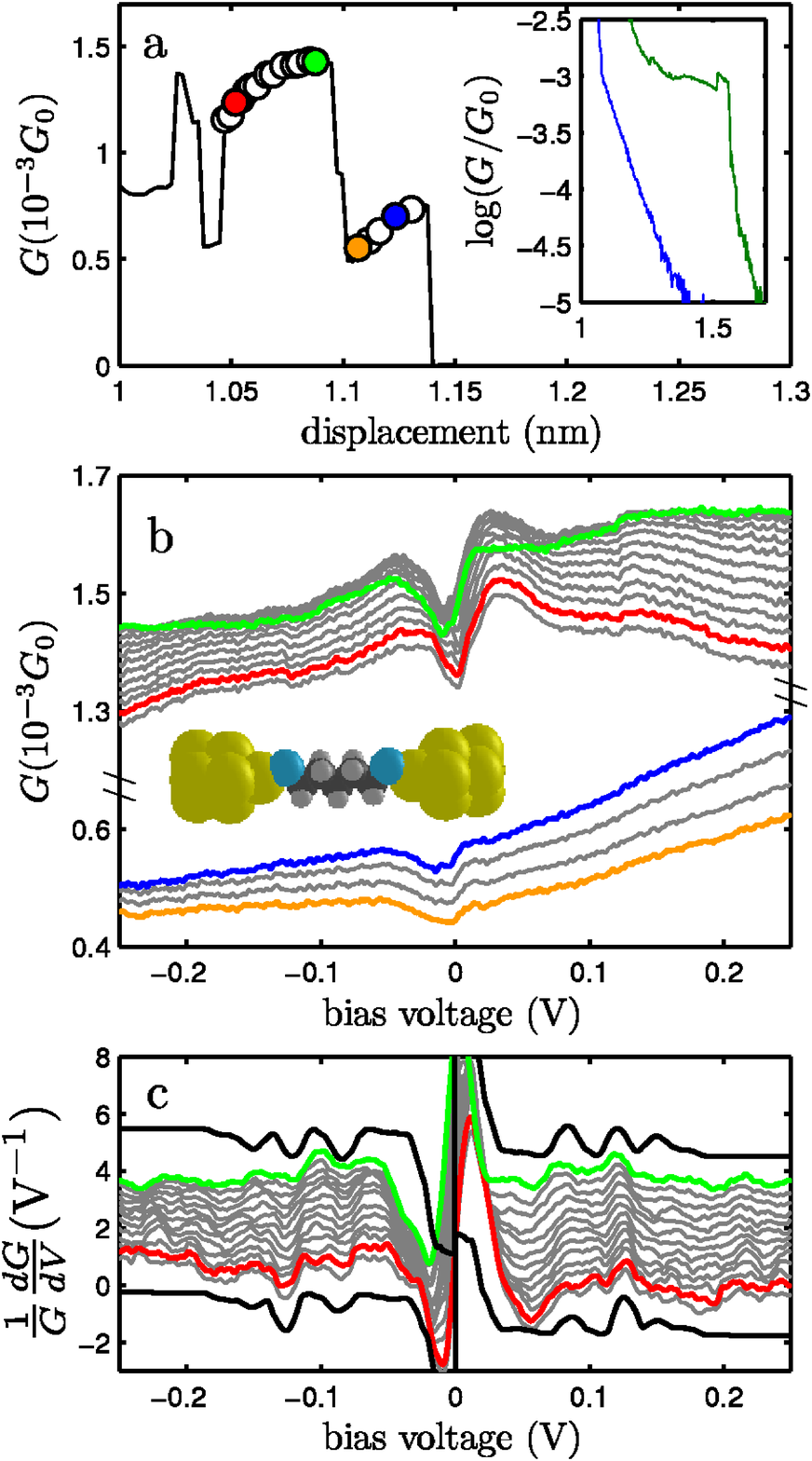}
\caption{(Color online)
(a) Selected conductance trace corresponding to the formation of a
stable molecular junction. Note the jump in conductance at a displacement
of about 1.1 nm, caused by the strain relaxation in the junction.
(Inset) Low-voltage conductance traces during electrode separation after breaking a gold
contact for a clean metal junction (blue trace)
and for a junction in which a molecule is present (green trace).
(b) Differential conductance vs voltage acquired at
the positions indicated by the circles in panel (a).
(c) Derivative of the differential conductance (normalized by the conductance),
$ G^{-1} d^{2}I/dV^2$, of the curves in panel (b).
For clarity, the experimental curves have been shifted upwards as the molecular junction is stretched.
The black thick curves are calculated spectra for two different electrode separations
that differ by 0.1 nm (top curve corresponds to a more stretched configuration
than bottom curve).}\vspace{-3mm}
\end{figure}

Single-molecule junctions are obtained by repeatedly forming and breaking the contact
between an STM tip and a gold substrate covered with pentanedithiol molecules.
The molecules were deposited on the substrate by immersion in a 1 mM solution 
of pentanedithiol molecules in toluene for 12 hours followed by washing and 
sonication in pure toluene, and drying in a
stream of helium gas. Immediately after that, the sample was transferred to 
the low temperature insert in helium gas atmosphere.
Before immersion
the substrate was treated with a piranha solution and flame annealed.
The tip consisted of a freshly cut 99.99\% purity gold wire. These deposition
conditions are typical for the formation of self-assembled monolayers (SAM),
and consequently we would expect a coverage close to a monolayer.
During the process of contact formation and breaking the conductance of the
junction is measured by recording the current  at a fixed bias voltage,
of typically 25 mV, following a procedure similar to the one
used for the study of Au atomic contacts \cite{untiedt1997prb}, single-atom
chains \cite{yanson1998nature,agrait2003phrep} and
single-propanedithiol junctions \cite{hihath2008nl}.

After rupture of a clean gold contact at low temperatures
the current at a constant tip-substrate bias
voltage decreases exponentially with a high apparent tunneling
barrier of 3--4 eV, which is characteristic of vacuum tunneling \cite{rubio2004prl}.
In contrast, the presence of the molecules on the substrate
gives rise to plateaus in the conductance, as
shown in the inset of Fig.~1(a). The lowest conductance plateau before
rupture is associated with a
single molecule suspended between substrate and tip
in a stable configuration \cite{xu2003science}.
Consistently with the deposition conditions,
the initial distribution of molecules on the substrate is quite homogeneous.
However, repeated junction
formation on the same spot resulted
in a depletion of molecules.
When one of these low conductance plateaus is detected, tip retraction is
stopped, and the current $I$, and differential conductance, $dI/dV$, are measured  as a function
of the applied bias voltage $V$. The differential conductance is measured using a lockin technique 
with an ac voltage modulation of $V_\mathrm{rms}=5$ mV.
The measurement is repeated in small incremental steps of
tip retraction, leading to the acquisition of data such as those summarized in Fig.~1.
The continuous parts of the conductance trace indicate regions where the junction
deforms elastically, while the sudden jumps in the conductance result from the
relaxation of the accumulated stress in the junction. Most likely these relaxations
are due to atomic rearrangements at the electrodes in the
immediate proximity of the junction, as demonstrated in atomic wires of gold
\cite{rubio2001prl}.

The differential conductance curves, as those shown  in Fig.~1(b),
show a continuous evolution within a conductance
plateau, while sudden changes in the overall shape of the curves are observed at
the conductance jumps.  As in the case of atomic contacts
\cite{ludoph2000prb,untiedt2000prb}, the large-scale features of the differential
conductance curves do not reflect the properties of the junction itself but are
a consequence of elastic scattering of the electrons in the vicinity of the
junction  and vary markedly from junction to junction.  Often the differential
conductance curves show a dip at low voltages. These so-called  zero-bias
anomalies are poorly understood and   typical of low temperature
junctions.  We will not give them further consideration in this work.

Careful inspection of the differential conductance curves in Fig.~1(b) shows
small conductance jumps of the order of one percent at certain voltages, the most
prominent of which occurs at $\pm 120$ mV. These sudden changes
in conductance are known signatures of inelastic scattering in the
junction \cite{agrait2003phrep}. At low temperatures the vibrational motion
of the molecule, which are almost completely frozen, can only be excited by the
passing electrons provided
the applied bias exceeds the quantum of a given vibration.
In the low-transmission, off-resonance case
this onset of inelastic scattering leads to enhanced conductance \cite{paulsson2005prb,delavega2006prb},
and hence to peaks (dips) in $d^2I/dV^2$ at positive (negative) bias polarity.
Fig.~1(c) shows the IETS spectra, calculated by taking the numerical derivative
of the measured differential conductance curves in Fig.~1(b). Indeed, we observe
antisymmetric features expected for the inelastic scattering processes.

\begin{figure}
\includegraphics[width=8 cm]{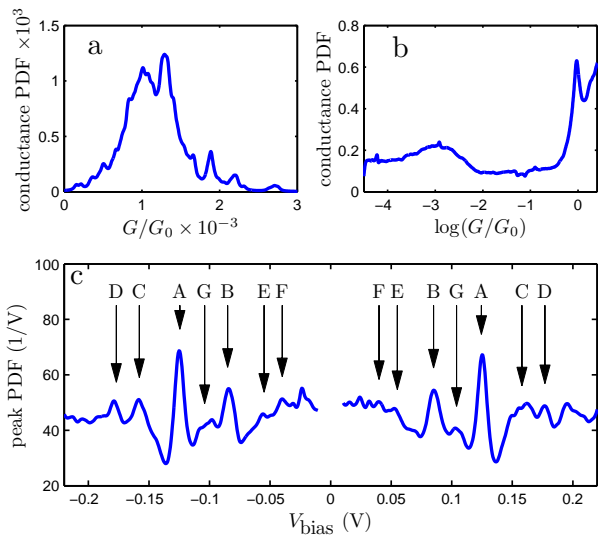}
\caption{(Color online) (a) Probability density function (PDF) of the low-bias conductance  
for 540 spectra in 33 different molecular junctions 
acquired at low temperatures (4.2 K). The low-bias conductance of a 
given spectra is the ratio $I/V$ at $V=25$ mV obtained from the simultaneously 
measured $I$ vs $V$ curve. In the PDF we have used $\epsilon = 3\times 10^{-5}$.
(b) PDF for the low-bias conductance measured at room
temperature in solution (green curve). 
(c) PDF of peaks (dips) in the IETS spectra at positive (negative) voltages.
Voltages around zero have been excluded due to the presence of zero-bias anomalies.
In the PDF we have used $\epsilon = 3$ mV.}
\vspace{-3mm}
\end{figure}

We have studied 33 different stable molecular junctions characterized by a well-defined 
conductance plateau at low conductance, obtaining a total of 540 IETS spectra from  
the measured differential conductance curves.  The probability density function (PDF) 
of the low-bias conductance, plotted in Fig.~2(a), shows that most of the 
mechanically stable junctions have a conductance of approximately
$10^{-3} G_0$. This value is consistent with the conductance peak in the PDF
observed at room temperature in solution shown in Fig.~2(b), and supports the
idea that transport through these molecular junctions takes place via tunneling.
Note that these two PDFs correspond to two different situations and are not 
directly comparable: in Fig.~2(b), the  PDF is obtained from all measured traces 
and includes all points of the trace while in Fig.~2(a)  the PDF includes a single point per spectra. 
We have chosen to use probability density (PDF) instead of the conventional histograms, so that 
the probability of finding a value in a certain range can be obtained by integration 
of the PDF in that range. To obtain the PDF, $f(x)$ from a set of experimental data, 
$x_i$ we use the standard definition, $f(x) = (1/N)\sum_i  \delta(x-x_i)$, where $N$ is 
the number of data, and we approximate the Dirac delta function to a Lorentzian function, 
$\delta(x) \approx \frac{\epsilon / \pi}{x^2+\epsilon^2}$, with an adequately small $\epsilon$.

In Fig.~2(c), we show the PDF of peaks (dips) in the IETS spectra appearing at positive 
(negative) voltages. All peaks above the noise level  are used to construct the PDF 
without further selection. We have normalized the PDF to the number of spectra and 
consequently this function represents the probability of finding a peak in a given 
interval of the bias voltage.
The symmetry of the features reflects the
antisymmetry of the IETS spectra. Some peaks are present in all spectra.
More precisely, the probability is unity for finding a
peak in the interval ($125\pm10$) mV (peak of type A),
as well as for finding one in ($85\pm13.5)$ mV (peak of type B).  These signals can be
observed directly in the individual spectra, cf.~Fig.~1.

During the measurement of the IETS spectra the  molecular junction is subject
to a varying stress: the junction is stretched up to the breaking point. The
effect of this stretching on the vibrational spectra is to shift the position of
peaks of types A and B to lower voltages (a maximum of 5-10 mV), i.e., to lower
frequencies, as shown in Fig.~3.  It is not possible to follow the evolution of the
other modes as the corresponding peaks are less well-defined. The observed
frequency redshift is similar to that
observed for atomic gold chains \cite{agrait2002prl}, where the inelastic
signal originates from a longitudinal mode that softens with stretching,
reflecting the weakening of the interatomic bonds \cite{frederiksen2004prl}.
The conductance jump at a displacement of 1.095 nm in Fig.~1(a)
results in a frequency jump in Fig.~3, demonstrating that its origin is in a
strain relaxation of the junction, probably due to an atomic rearrangement
in the electrodes \cite{rubio2001prl}.

\begin{figure}
\includegraphics[width=8 cm]{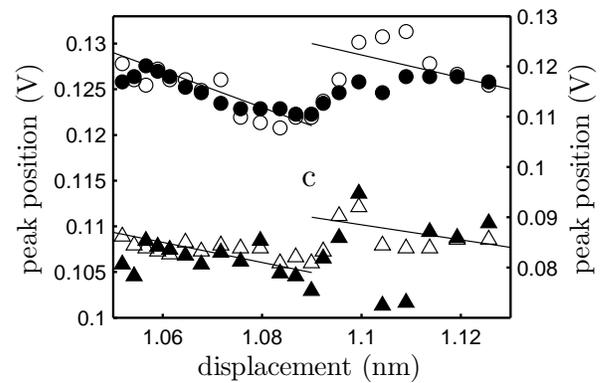}
\caption{
Peak position as a function of the tip retraction in Fig.~1.
Circles correspond to peak A around 125 mV (left $y$-axis)
and triangles to peak B around 80 mV (right $y$-axis). Filled (open) symbols
represent data obtained at positive (negative) voltage polarity.
The continuous lines are a guide to the eye.
}\vspace{-3mm}
\end{figure}

\begin{figure*}[t!]
\includegraphics[width=16cm]{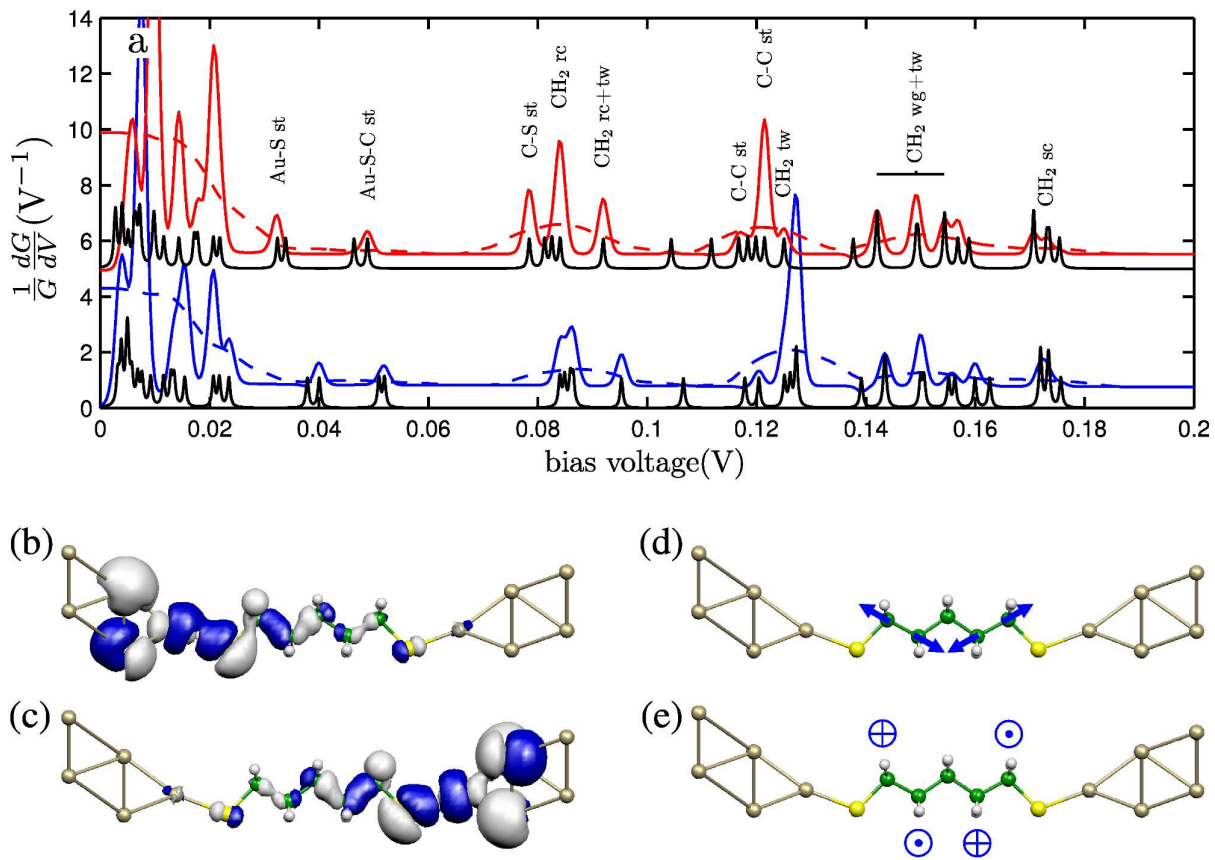}
\caption{(Color online) (a) Calculated IETS spectra of pentanedithiol for two different
electrode separations that differ by 0.1 nm (top-red more stretched than bottom-blue).
The dashed curves are rounded using a smoothing filter equivalent to
that used in the processing of the experimental data.
The
corresponding black curves (vertical axis scale arbitrary) represent the density
of vibrational modes for these two configurations. The different peaks are labeled
according to the character of the mode (st=stretch, rc=rock, wg=wag, tw=twist,
sc=scissor).
(b),(c) Eigenchannel scattering states \cite{paulsson2007prb} corresponding to
electrons originating in the left and right electrode, respectively. The visualizations
represent isosurfaces of the real part of the wave functions (with sign). (d),(e)
Illustration of an active (d) and a silent (e) mode from the spectrum for the less
stretched configuration.}\vspace{-3mm}
\end{figure*}

In order to understand how the peak positions and amplitudes in the IETS spectra
relate to the various vibrational modes of the molecular configuration
and their dependence on stretching,
we have performed first-principles transport calculations based on non-equilibrium
Green's functions techniques \cite{brandbyge2002prb,frederiksen2007prb}.  Our simulations attempt to mimic the
experimental situation with the molecule suspended between two metal electrodes, as shown in
Fig.~4, assuming that the S atoms are bound to Au adatoms on
Au(111) surfaces \cite{nota}.
The calculated zero-bias conductance of the pentanedithiol junctions is of the order
0.005 G$_0$ per molecule, i.e., about a factor of 5 higher than the experimental value.

Fig.~4(a) shows the calculated vibrational spectrum for a pentanedithiol
between two Au electrodes for two different electrode separations. The spectra are quite similar
to those calculated for octanedithiol junctions \cite{paulsson2009nl}.
We observe that there are active and inactive
modes due to approximate selection (propensity) rules \cite{paulsson2008prl}.
The strength of the IETS signal can be related to a matrix element involving three
factors: initial electron state, vibrational deformation potential, and final
electron state. Therefore, propensity rules can be understood from symmetry
considerations. For the pentanedithiol junction an eigenchannel analysis shows
that the IETS can be rationalized by just considering inelastic scattering
between the two scattering states belonging to the most transmitting
eigenchannel. These states, shown in Fig.~4(b) and (c) for electrons
originating in the left and right electrode, respectively, possess a $\sigma$-type
symmetry and can therefore only couple via modes with longitudinal character.
For instance, we get a large signal from the C--C stretch mode
($\hbar\omega=127$ meV) illustrated in Fig.~4(d), while scattering is suppressed for the
essentially transverse mode ($\hbar\omega=106$ meV) illustrated in Fig.~4(e).

For comparison with the experimental curves the rounding introduced by the ac
modulation used to measured the differential conductance
has also been included in the theory curves, shown with dashed lines in Fig.~4(a)
and the  top and bottom curves (thick black traces) in Fig.~1(c).
As a
result of this rounding the individual peaks cannot be resolved.
Fig.~1(c) shows the excellent agreement between experiment and theory, even the magnitude of
the inelastic signals is quantitatively reproduced.
All the peaks appearing in the PDF in Fig.~2(c) can be identified with some group of peaks in
Fig. 4(a):
peak A corresponds to a C--C stretch with some contribution from CH$_2$ twist; B to
C--S stretch with some CH$_2$ rock; C to CH$_2$ wag and twist; D to CH$_2$ scissor;
E to Au--S--C stretch; F to Au--S stretch; and G to CH$_2$ rock and twist.
The fact that
the calculated peaks have the same intensities as the measured peaks [see Fig.~1(c)]
allows us to draw an important conclusion: the electrons are indeed tunneling
through a pentanedithiol molecule suspended between the electrodes.

Comparison with the experiment also requires to take into consideration that the
molecules are in different states of strain. We have calculated the spectra for
several different elongations of the molecule to determine the strain dependence
of the modes. In this way we can characterize the C--C (C--S) mode, i.e., peak A (B),
by a shift of $-5.0$ ($-2.9$) meV/{\AA}, consistent with the experimental results.
This is a strong indication 
that the junction consists of a single molecule. Indeed, if there
were more molecules in the junction contributing to the transport, they would
certainly be in different states of strain due to irregularities of the
electrodes and, thus, preclude the observation of clear frequency shifts.

In conclusion, our results show that combining first-principles transport calculations and 
high resolution low temperature IETS it is possible to characterize the configuration 
of single-molecule junctions. The agreement between experiment and theory is excellent and, 
in addition to identifying the presence of pentanedithiol molecules, allows us
to extract structural information of the molecular junction, e.g., its state of strain. 
Indeed, the IETS signal indicates that the electrons are passing through the backbone of 
the molecule and that the it is bound to the electrodes by the thiol groups. 

TF, DSP and AA acknowledge stimulating discussions with Magnus Paulsson.
This work has been supported by the Spanish MICINN (MAT2008-01735,
FIS2007-6671-C02-00 and MAT2007-62732,
and Consolider-Ingenio 2010 CSD2007-0010), the CAM (``CITECNOMIK'' P-ESP-000337-0505),
the UPV/EHU (IT-366-07), and the Basque Depto.~de Industria and
the Diputaci\'on Foral de Guipuzcoa (``ETORTEK''),
the EC (STREP ``SURFMOF'' NMP4-CT-2006-032109, and FP7 ITN ``FUNMOLS'' 212942).
TF acknowledges support from the Danish FNU (Grant No. 272-07-0114).

\end{document}